\pdfoutput=1
\documentclass{emulateapj}
\usepackage{amssymb,fancyhdr,lastpage,layout,bold-extra,floatrow,calc,wrapfig}
\usepackage[backref,breaklinks,colorlinks,allcolors=blue]{hyperref}
\usepackage{graphicx}

\usepackage{natbib}
\usepackage{aas_macros}
\usepackage[section] {placeins}
\usepackage{subfigure}
\usepackage{xspace}
\usepackage{float}
\usepackage{color}

\usepackage[scaled]{helvet}

\usepackage[T1]{fontenc}

\newcommand{\mach}{$\mathcal{M}$}
\def\h2{${\rm\,H_2}$}

\usepackage{color}

\begin{document}

\title{The Impact of Unresolved Turbulence on the Escape Fraction of Lyman Continuum Photons}
\author{Mohammadtaher Safarzadeh\altaffilmark{1} and Evan Scannapieco\altaffilmark{1}}

\footnotetext[1]{School of Earth and Space Exploration, Arizona State University, Tempe, AZ 85287-1404, USA;\\ Email: mts@asu.edu}

\begin{abstract} 
We investigate the relation between the turbulent Mach number (\mach) and the escape fraction of Lyman continuum photons ($f_{\rm esc}$) in high-redshift galaxies. 
Approximating the turbulence as isothermal and isotropic, we show that the increase in the variance in column densities from $\mathcal{M}=1$ to $\mathcal{M}=10$ causes  $f_{\rm esc}$ to increase by $\approx 25$\%, and the increase from $\mathcal{M}=1$ to $\mathcal{M}=20$ causes $f_{\rm esc}$ to increases by $\approx 50$\% for a medium with opacity $\tau\approx1$. At a fixed Mach number, the correction factor for escape fraction relative to a constant column density case scales exponentially with the opacity in the cell, which has a large impact for simulated star forming regions.
Furthermore, in simulations of isotropic turbulence with full atomic/ionic cooling and chemistry, the fraction of HI drops by a factor of $\approx 2.5$ at $\mathcal{M}\approx10$ even when the mean temperature is $\approx5\times10^3 K$. If turbulence is unresolved, these effects together enhance $f_{\rm esc}$ by a factor $>3$ at Mach numbers above 10. Such Mach numbers are common at high-redshifts where vigorous turbulence is driven by supernovae, gravitational instabilities, and merger activity, as shown both by numerical simulations and observations.  These results, if implemented in the current hydrodynamical cosmological simulations to account for unresolved turbulence, can boost the theoretical predictions of the Lyman Continuum photon escape fraction and further constrain the sources of reionization.
\end{abstract}
 
\section{Introduction}

Observations of high-redshift quasars indicate that the universe was reionized by $z=6$ \citep{Fan:2002dh,Fan:2006bk,Becker:2007fj,Mortlock:2011ky}. Similarly, Planck satellite measurements constrain the  reionization optical depth to be $\tau=0.066\pm0.016$, corresponding to a redshift of $z\approx8.8$ \citep{Collaboration:2016bk}. However, there is a debate as to whether dwarf galaxies or quasars provided the ionizing photons that caused this transition \citep{Volonteri:2009ha,Grissom:2014gz}.  An escape fraction of ionizing photons, $f_{\rm esc}$, above $10\%$ may be necessary to explain reionization by dwarf galaxies alone \citep{Wise:2009fn,Kim:2013dh}, while still higher levels of $z>9$ ionizing photons are necessary to explain the observed electron optical depth observed by Planck. 

Although the correlation between $f_{\rm esc},$ halo mass, and redshift has been studied by various groups, the results are often inconsistent,  with a wide range of $f_{\rm esc}$ values reported in the literature \citep[see][and references therein]{Xu:2016wk}. \citet{Gnedin:2008ib}, for example, find that $f_{\rm esc}$ increases in galaxies with higher star formation rates because in these galaxies young stars are found at the edge of the HI disks and are therefore not enshrouded in the disk's gas. However, they also find that $f_{\rm esc}$ is less than 0.01 for systems with total masses below  $5 \times 10^{10} M_\odot,$ a number that is very difficult to reconcile with  reionization measurements \citep{Gnedin:2008in}. Similarly, for high-redshift systems, \citet{Wood:2000ie} argued that higher disk densities at high redshifts would lead to a drop in the escape fraction, with $f_{\rm esc} \approx0.01$ for sources at $z\approx10.$  On the other hand, recent high-resolution zoom simulations by \citet{Kimm:2014gv} suggest that runaway OB stars and supernovae may lead to escape fractions as high as $f_{\rm esc}\approx 14\%$.

A key question regarding these estimates is how unresolved turbulent structures impact the escape fraction of LyC photons.  Accounting for these structures would mean capturing the underlying probability distribution function of gas density and correcting the  column density distribution of gas in the host galaxies containing massive stars. To achieve this one would need to resolve the driving scale of the turbulence ($L_{\rm drive}$) by at least 50 elements \citep[e.g.][]{Gray:2015gz}. 

Estimating $L_{\rm drive}$ is not an easy task, as it could be set by many plausible mechanisms, including supernova feedback, gravitational instability, the infall of gas clumps, and thermal instabilities \citep{Wise:2007fn,Wise:2008fl,Green:2010kv,Krumholz:2016ji}. For example, \citet{Chepurnov:2015jt} measure a driving scale of 2.3 kpc based on the velocity power spectrum of the Small Magellanic Cloud (SMC)  which implies that large-scale galaxy-galaxy interactions drive the turbulence. However, \citet{Martizzi:2016de} find $L_{\rm drive}\approx100$ pc in their supernova driven vertically stratified medium, consistent with the observations of \citet{Dib:2009dg} which find a $L_{\rm drive} \approx 100-800$ pc based on the orientation of molecular clouds with respect to the galactic plane. Furthermore, it is expected that the driving scale decreases at higher redshifts, because galaxies become smaller \citep{Ferguson:2004dt} and clumpy in their morphology \citep{Guo:2015dr}. Therefore it would be rather difficult to claim that turbulence is resolved in simulations of high-redshift galaxy formation. 

Failing to resolve this structure would smear out the probability distribution function (PDF) of the supersonic turbulent gas, which would otherwise  approximately follow a log-normal distribution with a variance that is a function of the turbulent Mach number \citep{VazquezSemadeni:1994fm,Ostriker:2001ky,Hopkins:2013hn,Nolan:2015jo}. Conservation of mass implies that at higher \mach, the peak of the density PDF shifts to lower values and  the PDF becomes broader. This increase in the variance of the PDF increases the relative fraction of very low column densities, which leads to low column density sight lines that are ideal for the escape of ionizing photons. 

Furthermore, recent simulations of \emph{non-}isothermal isotropic turbulence, show that even at temperatures as low as T$\approx5\times10^3K$, the HI fraction drops by a factor of $\approx2.5$ for the studied range of Mach number $1<\mathcal{M}<12$ \citep{Gray:2016ex} and more so when the background ultraviolet light is taken into account. Like the increase in the variance of the column depth described above,  this drop in the HI mass fraction would increase the probability that a photon would escape a higher Mach number medium due to unresolved turbulence.  

In this letter, we bring these two facts into attention and argue that their combined effect is sufficient to increase current estimates of the escape  by a factor three or more in cases in which the average Mach number of the medium is above \mach $\approx7.$ The structure of this work is as follows:  In \S2 we describe a simple model of how the escape fraction scales with  Mach number, in \S3 we describe how this scaling is likely to affect current estimates of $f_{\rm esc}$ as a function of redshift, and in \S4 we give conclusions.

\section{Modeling a Turbulent Medium}

The density distribution in a non-magnetized supersonic box is well approximated by a log-normal distribution where
\begin{equation}
\sigma^2_{\rm ln\rho} = \ln (1+b^2 \mathcal{M}^2),
\end{equation}
\citep{VazquezSemadeni:1994fm,Ostriker:2001ky}. The PDF of column densities ($\Sigma$) at the driving scale
is also well-approximated by a lognormal distribution \citep{Ostriker:2001ky, Federrath:2010ef,Burkhart:2015jz}:
\begin{equation}
p(x,\mathcal{M})=\frac{1}{(2\pi\sigma_{\ln \Sigma}^2)^{1/2}}\exp\left[-\frac{(x-\overline{x})^2}{2\sigma_{\ln \Sigma}^2}\right],
\end{equation}
where $x \equiv \ln (\Sigma/\langle\Sigma\rangle)$. Conservation of mass requires that the mean $\overline{x}$ and 
dispersion $\sigma_{\ln\Sigma}$ in $p(x)$  be related by $\overline{x}=-\sigma_{\ln\Sigma}^2/2$. 
By assuming a form of power spectrum for density fluctuations, the variance in the two-dimensional projected column density is related to the variance in  the three-dimensional density field \citep{Brunt:2010eh}. In this case \citet{Thompson:2016gz} find the dispersion of the logarithm of column density to be 
\begin{equation}
\sigma_{\rm ln\Sigma}\approx \ln (1+ R b^2\mathcal{M}),
\end{equation}
where
\begin{equation}
R = \frac{1}{2}\left(\frac{3-\alpha}{2-\alpha}\right)\left[\frac{1 - \mathcal{M}^{2(2-\alpha)}}{1 - \mathcal{M}^{2(3-\alpha)}}\right],
\end{equation}
and $\alpha$ is the slope of the turbulent power spectrum $P(k)\propto k^{-\alpha}$ for $1 \leq k \leq \mathcal{M}^2$ and zero otherwise. Here $k$ is 
in normalized units such that $k_x=1$ corresponds to a mode with with wavelength $\lambda= 2 L_{\rm drive}.$ 
\citet{Federrath:2008ey} showed $b=1$ for purely compressive forcing ($\Delta\times F=0$) and $b=1/3$ for purely solenoidal ($\Delta.F=0$).
\citet{Gray:2015gz} found $b=0.53$ best fits their simulation result where they relax the assumption of isothermal turbulence and 
model the cooling and atomic chemistry of the gas in a supersonic turbulent box.

While both HI and dust can absorb the LyC photons, dust is found to have a negligible impact on the final escape fraction of photons in the presence of neutral Hydrogen \citep{Gnedin:2008ib}. This is because the dust opacity is in general much less than HI ($\tau_{\rm dust}<<\tau_{\rm HI}$) for ionizing photons  and for lines-of-sight where $\tau_{\rm dust}>1$, the hydrogen opacity is already so large that it dominates.  
To model the effect of increasing the Mach number on the escape fraction of LyC photons, we therefore take the distribution of HI column densities, multiply by the opacity of each column which is directly proportional to the column density and integrate over all the column densities:
\begin{equation}
 Esc(\mathcal{M},\tau_{\rm mean})=\int_{-\infty}^{{+\infty}}  p(x,\mathcal{M}) e^{- \tau_{\nu}} dx,
\end{equation}
where $\tau_{\nu}= \kappa_{\nu} \Sigma=\kappa_{\nu} \langle\Sigma\rangle e^x$. $\kappa_{\nu}$ is the hydrogen opacity per unit mass in ${\rm cm}^2/{\rm gr}$ at frequency ($\nu$). The escape fraction is a function of both \mach\, and $\tau_{\rm mean}$ (the product of $\kappa_{\nu} \langle\Sigma\rangle$).
We compute the integral as a function of these two parameters and report the relative enhancement of escape fraction as compared to the case assuming a constant column density of $\langle\Sigma\rangle$. Only a specific range in $x$ makes most of the contribution to the integral, because the PDF drops at either too high or too low column densities. Moreover, high column densities suffer from large absorption which makes them irrelevant for contributing to the escape fraction. This point is illustrated in Figure 1.

It should be noted that in our calculations we have assumed the scale height ($H$) of the galaxy in the question is comparable or less than 
the driving scale of turbulence. If, on the other hand, the scale height of the system is larger than the driving scale ($H>L_{\rm drive}$), then the PDF of column densities would be the product of of $N$ lognormal PDFs characterized by $\sigma_{\rm ln\Sigma}=\sqrt{N}\sigma^{\rm drive}_{\rm ln\Sigma}$ and 
$\langle\Sigma\rangle=N\langle\Sigma\rangle^{\rm drive}$ where $N=H/L_{\rm drive}$.

Figure 1 shows the function $p(x)$ for three different Mach numbers as well as the normalized escape fraction computed for different ranges of  the lower limit on $x$. The lower panel plot indicates to what lower limits in column densities one needs to integrate, and therefore resolve in the simulations, in order 
to properly measure the escape fraction of ionizing photons.  At higher Mach numbers, we would need to integrate to lower values in $x$ to compute the total escape fraction. The lower limit that would suffice is $x\approx-3$ at $\mathcal{M}=20.$ 

\begin{figure}
\resizebox{3.2in}{!}{\includegraphics{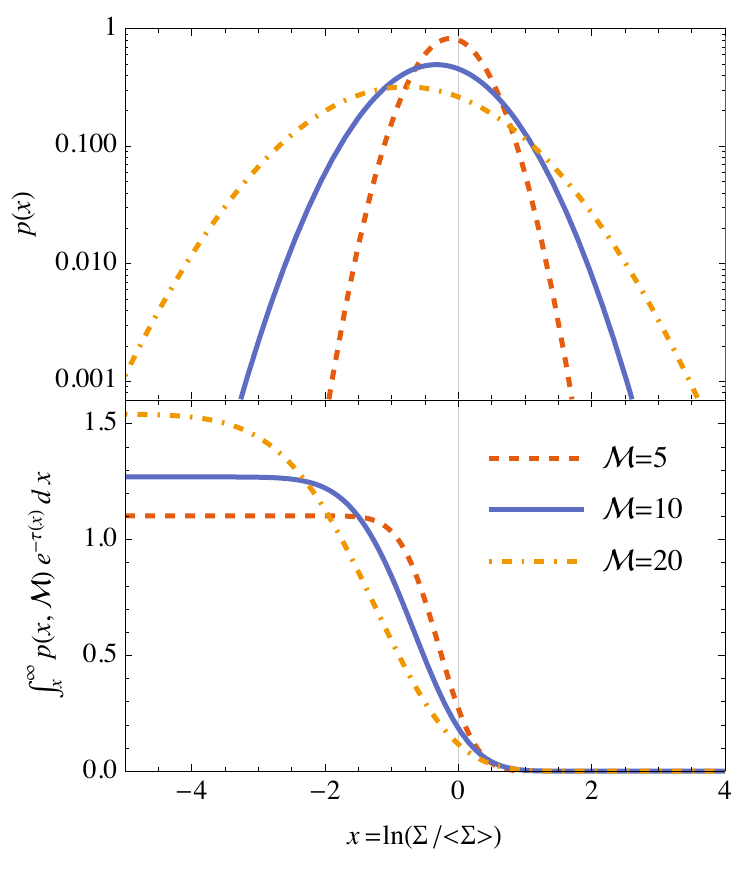}}
\caption{{\em Top:} The qualitative behavior of $p(x),$  the probability density function of ln column density,  computed for three different representative Mach numbers. {\em Bottom:}  The impact of very low column densities on the escape fraction relative relative to the case assuming a constant column density of $\langle\Sigma\rangle$ for a grid cell. We have considered $b=1/2$ for all the curves and assumed a turbulent power spectrum with $p(k)=k^{-\alpha}$ with $\alpha=2.5$. 
We have assumed $\kappa_{\nu} \langle\Sigma\rangle=1$ in this calculation.}
\end{figure}

Figure 2 shows $Esc(\mathcal{M})$ normalized at $Esc(\mathcal{M}$=1) for different values of $b$: 0.33,1 and 0.53, which correspond to an
isothermal medium with solenoidal forcing, an isothermal medium with compressive forcing, and a non-isothermal medium with solenoidal forcing, respectively. The most important case is the result with $b=0.53$ which is what is found when the assumption of isothermality is relaxed.  Note that this is very similar to $b^2=1/4$ implemented in other studies \citep[][and references therein]{Thompson:2016gz}. In this case, we see a smooth rise in the relative escape fraction with increasing Mach number,  with a 25\% increase in escape fraction at $\mathcal{M}\approx10$ and a 50\% increase at $\mathcal{M}\approx20.$ Turbulence can be partially captured in a simulation and a direct implementation of our results would account for the \emph{unresolved} turbulence in a simulation grid cell. The unresolved turbulence could be measured based on the shear forces on a grid cell as in the Smagorinsky model \citep{Smagorinsky:1963}.

\begin{figure}
\resizebox{3.2in}{!}{\includegraphics{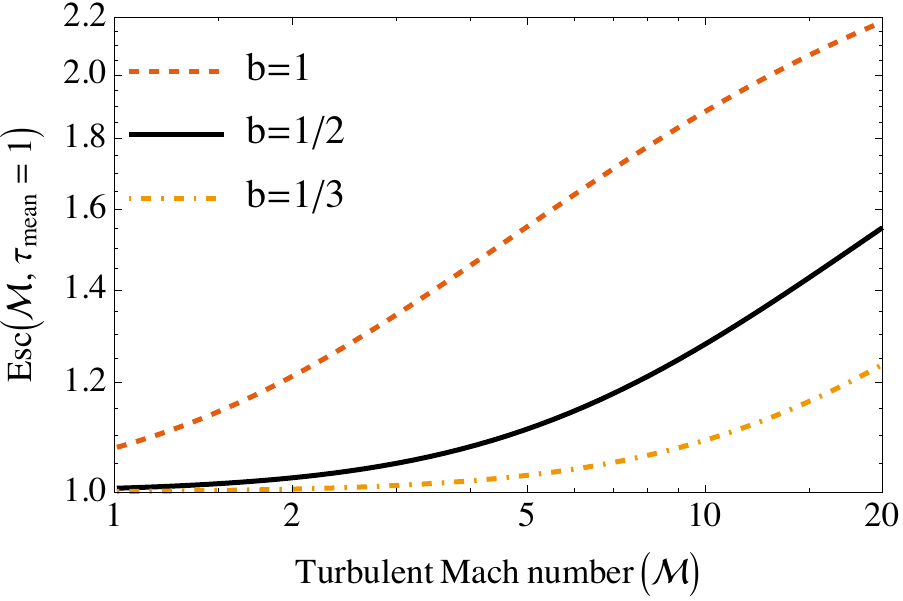}}
\caption{ The enhancement of the escape fraction of LyC photons as a function of \mach\, relative to fixed column density of $\langle\Sigma\rangle$. The three curves correspond to three values of $b:$ 0.33,1 and 0.53, which correspond to an isothermal medium with solenoidal forcing, an isothermal medium with compressive forcing, and a non-isothermal medium with solenoidal forcing, respectively. The escape fraction is modeled based on Eq.\ (5), assuming a turbulent power spectrum with $p(k)=k^{-\alpha}$ and $\alpha=2.5$. The escape fraction is increased by 25\% at $\mathcal{M}\approx10$ and by 50\% at $\mathcal{M}\approx20$ for the case of $b=1/2$ which is the case when detailed chemistry is calculated for the turbulent gas.  This calculation does not take into
account the reduction of the HI fraction in turbulent media. We have assumed $\kappa_{\nu} \langle\Sigma\rangle=1$ in this calculation.}
\end{figure}

Figure 3 shows the effect of opacity on the escape fraction. In highly opaque cells, introducing the column density PDF of a 
supersonic turbulence make a large impact on the escape fraction of photons relative to a constant density assumption. This potentially have 
a large impact on the escape of ionizing photons from star forming regions in a simulation in that star formation takes place in more opaque regions than the average opacity of
the galaxy.

\begin{figure}
\resizebox{3.2in}{!}{\includegraphics{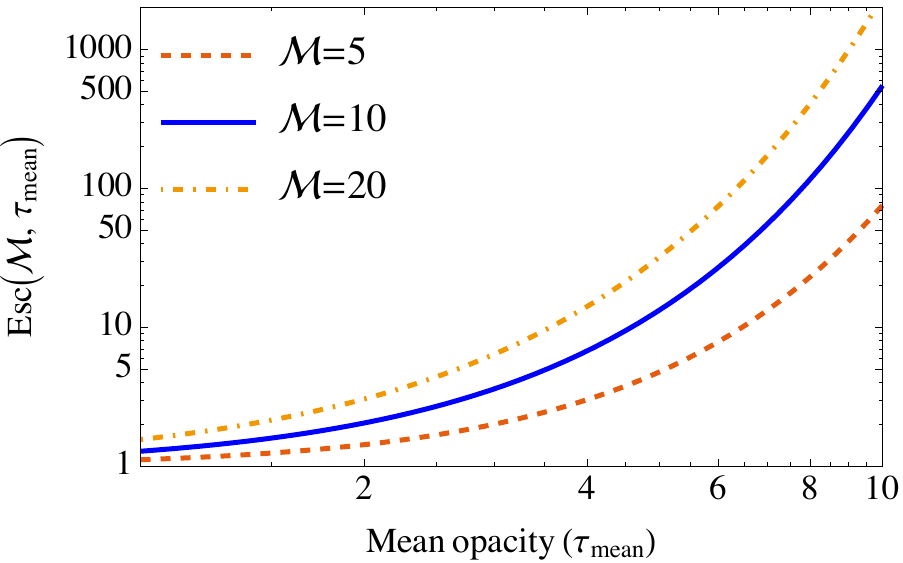}}
\caption{The enhancement of the escape fraction of LyC photons as a function of $\kappa_{\nu} \langle\Sigma\rangle$ (mean opacity) relative to fixed column density of $\langle\Sigma\rangle$. The three curves correspond to three values of \mach: 5,10 and 20 and we have considered $b=1/2$ for all the curves. The  escape fraction is modeled based on Eq.\ (5), assuming a turbulent power spectrum with $p(k)=k^{-\alpha}$ and $\alpha=2.5$. Resolving turbulence becomes more crucial for more opaque cells in a simulation and this can potentially have a large impact on the escape of ionizing photons in star forming cells which are generally located in highly opaque regions.}
\end{figure}

The calculations so far assume that the abundance of neutral hydrogen does not change with the increase of turbulent Mach number. 
However, recent non-isothermal, isotropic simulations show that even when the average temperature is $5\times10^3$ K, the HI fraction 
can drop below 50\% for a wide range of $1<\mathcal{M}<12$ \citep{Gray:2016ex}. Note that the density variance-Mach number relation in this case is similar to what we have assumed in our calculations \citep{Gray:2015gz,Gray:2016ex}.

Other studies have considered the impact of self-gravity and thermal energy from supernovae  on the turbulent density distribution.
 \citet{Slyz:2005dk}, like \citet{Gray:2015gz}, found that the density PDF is  lognormal when one includes cooling and turbulent driving. They also showed that the addition of self-gravity causes the PDF to exhibit a high-density tail, corresponding to collapsing gravitationally bound structures \citep{Klessen:2000df,Federrath:2013ip}. However this has little effect on the   low density end that determines $f_{\rm esc}$. When thermal feedback is included in their isotropic simulations, however, the PDF becomes bimodal as most of the volume becomes heated to high temperatures. Although, in a real galaxy, much of this hot gas is likely to be vented away in an outflow, its presence will nevertheless cause $f_{\rm esc}$ to increase even more than in our estimates above. Thus while various pieces of physics can affect the density PDF in a supersonic turbulent gas, the results show these changes will generally enhance the low density  part of the density PDF and therefore will further enhance the escape fraction. 
 
 
\section{Turbulent Mach Number at High Redshifts}

Regardless of the detailed physics, the efficacy of small-scale turbulent structures in facilitating reionization will depend on turbulent Mach numbers being large in high-redshift galaxies. Yet there are several theoretical and observation clues that this may indeed be the case.
From a theoretical point of view, \citet{Wise:2008fl} found that turbulent Mach number rises to $\mathcal{M}\approx2-4$ in simulations of galaxies at high redshifts, 
when the halos becomes gravitationally unstable and start to collapse.
\citet{Greif:2008ke} show that inflow of cold gas along high-redshift cosmic filaments is supersonic, leading to a high level of $\mathcal{M}\approx1-5$ turbulence in accreting galaxies.  \citet{Sur:2016ke} discuss the turbulent velocity dispersions expected to be caused by gravitational instabilities in rotationally supported structures, showing they are likely to be much greater at high redshifts.

From an observational standpoint, at high redshifts, smaller sizes and higher temperatures can  create extreme environments that lead to highly-supersonic structures \citep{Chabrier:2014cu} with 10<\mach < 100. Disk settlement happens at redshifts $z<1.2$ \citep{Kassin:2012kz} and the structure of galaxies 
at higher redshifts is clumpy in nature \citep{Guo:2012jn,Mandelker:2014be,Moody:2014gk,Guo:2015dr,Mandelker:2016db} with line-of-sight velocity dispersions typically five times higher than in the local universe \citep{Green:2010kv}. 
These clumpy structures indicate that disks at high redshifts are not stable, which leads to a highly-turbulent interstellar medium (ISM).

A highly-turbulent ISM is also observed at $z\approx2.3$ in the lensed star forming galaxy SMM J2135-0102 \citep{Swinbank:2011df}. The measured mid-plane hydrostatic pressure of SMM J2135-0102 is estimated to be $P_{\rm tot}/k_B\approx(2\pm 1)\times10^7 \, {\rm K} \,  {\rm cm}^{-3}$ which is $\approx1000\times$ higher than what is found for the Milky Way and only comparable to local ultra luminous infrared galaxies  \citep{Downes:1998co}. In fact \citet{Swinbank:2011df} find that supersonic turbulence is dominant on all scales down to $\approx100\times$ smaller than kinematically quiescent ISM of Milky Way and a supersonic turbulent theory can explain such observations \citep{Rathborne:2014ky}. Strong lensing  studies of the ISM of high-redshift galaxies appear to be the path forward to test the theories of turbulent structure at redshifts relevant for reionization.

\section{Conclusions}

We have shown that the escape probability of LyC photons increases significantly with Mach number in supersonic turbulent gas.  This is due to two reasons.  First,  at higher Mach numbers, the column density PDF becomes broader, characterized by the relation between the dispersion of the PDF and the Mach number of the flow.
This leads to more sightlines with very low opacities, which are the candidates for the escape of ionizing photons. Second, simulations of  supersonic turbulent flow which track the detailed abundance of the elements through chemical networks show a drop within a factor of 2-3 in the fraction of neutral Hydrogen at $\mathcal{M}\approx10$ even at temperatures $\approx5\times10^3$ K. These two effects combined can enhance the escape fraction of LyC photons by a factor of more than 3 in flows with $\mathcal{M}\approx10$, values expected to be typical in high-redshift galaxies.

The impact of resolving the density PDF on the escape fraction increases for the regions with high opacity (Figure 3).
However, it should be noted that star formation happens in dense regions and therefore the sources of ionization do not experience the average column density
of the galaxy they reside in. If a star is born in a high density region, an estimate of the Eddy turnover time scale given the unresolved turbulent velocity and cell size
($t_{\rm Eddy}\approx10 (\frac{\Delta(x)}{100\,{\rm pc}})(\frac{10\,{\rm{km/s} } }{v_{\rm turb}}) {\rm Myr}$) should be compared with the lifetime of the star ($t_{\rm age}\sim \rm{Myr}$). In the case $t_{\rm Eddy}>>t_{\rm age}$ then it would not be physical to replace the column density of the cell
with its equivalent supersonic turbulence box. The turbulent velocity and cell size are related through the assumed supersonic power spectrum.

Currently the detailed chemistry of species in an isotropic turbulent box is carried out up to \mach $\approx 10$. 
It can be that at higher Mach numbers, we might see an even more dramatic drop in the neutral Hydrogen fraction which would enhance the escape of LyC photons. Separately, a more careful study of the driving scale of turbulence in high-redshift galaxies needs to be conducted  to determine the minimum spatial resolution needed  to capture the turbulent structure of their ISM and its impact on the escape fraction of Lyman continuum photons.

\acknowledgements
We are thankful to Romeel Dave and Phil Hopkins for insightful conversations. 
This work was supported by the National Science Foundation under grant AST14-07835 and by NASA under theory grant NNX15AK82G. 

\bibliographystyle{apj}

\end{document}